 \documentclass{aa}
  \usepackage{psfig}

\newcommand{\HP}{{\sc hipparcos}}
\newcommand{\C}{Cepheids}

\begin{document}

\thesaurus{04(08.22.4, 08.04.1, 11.13.1 )} 
\title{The absolute magnitudes of RR Lyrae stars from {\sc hipparcos}
parallaxes\thanks{Based on data from the ESA Hipparcos astrometry
satellite.}  }

\author{M.A.T. Groenewegen \inst{1} \and M. Salaris \inst{2,1} }

\offprints{Martin Groenewegen (groen@mpa-garching.mpg.de)}

\institute{ Max-Planck Institut f\"ur Astrophysik,
Karl-Schwarzschild-Stra{\ss}e 1, D-85740 Garching, Germany
(groen@mpa-garching.mpg.de) \and 
Astrophysics Research Institute, Liverpool John Moores University,
Twelve Quays House, Egerton Wharf, Birkenhead CH41 1LD, UK
(ms@staru1.livjm.ac.uk)
}

\date{received,  accepted}

\maketitle

\begin{abstract}

Using the method of ``reduced parallaxes'' for the Halo RR Lyrae stars in
the \HP\ catalogue we derive a zero point of 0.77 $\pm$ 0.26 mag for an
assumed slope of 0.18 in the $M_{\rm V}$-[Fe/H] relation.  This is
0.28 magnitude brighter than the value Fernley et al. (1998a) derived
by employing the method of statistical parallax for the {\it
identical} sample and using the same slope. 

We point out that a similar difference exists between the ``reduced
parallaxes'' method and the statistical parallax method for the
Cepheids in the \HP\ catalogue.

We also determine the zero point for the $M_{\rm K}$-$\log
P_{0}$ relation, and obtain a value of $-1.16$ $\pm$ 0.27 mag (for a
slope of $-2.33$). The distance moduli to the \HP\ RR Lyrae stars 
derived from the two relations agree well.




The derived distance scale is in good agreement with the results from
the Main Sequence fitting distances of Galactic globular clusters and
with the results of theoretical Horizontal Branch models, and implies
a distance modulus to the LMC of 18.61 $\pm$ 0.28 mag.

\vspace{-1mm}

\keywords{RR Lyrae - Stars: distances - Magellanic Clouds}

\vspace{-2mm}

\end{abstract}

\vspace{-3mm}

\section{Introduction}

RR Lyrae stars are fundamental standard candles and the accurate
determination of their absolute luminosity has a wide range of
applications, including the derivation of the Hubble constant and the
determination of globular clusters ages.  The results of the \HP\
mission allow in principle a calibration of this luminosity, based on
the parallaxes and proper motions.

Fernley et al. (1998a, hereafter F98) did that by employing the method
of statistical parallax on a sample of 84 RR Lyrae stars (out of the
144 they considered) with [Fe/H] $\leq -1.3$. Combining the
statistical parallax result with the absolute magnitude of RR Lyrae
itself, computed without applying any Lutz-Kelker (LK) type correction
(see Lutz \& Kelker 1973, Turon Lacarrieu \& Cr\'ez\'e 1977, Koen
1992, Oudmaijer et al. 1998), they derived a zero point of 1.05 $\pm$
0.15 mag for the $M_{\rm V}$-[Fe/H] relation, by assuming a slope of
0.18 $\pm$ 0.03 (Fernley et al. 1998b).  Tsujimoto et al. (1998,
hereafter T98) used the statistical parallax method, a maximum
likelihood technique and the derived $M_{\rm V}$ of the star RR Lyrae
(with LK correction included) for deriving a combined final value
$M_{\rm V}$ = 0.6-0.7 mag at [Fe/H] = $-1.6$.  Luri et al. (1998,
hereafter L98) applied a maximum-likelihood method that takes all
available data into account, including parallaxes, proper motions and
radial velocities, considering the sample of 144 RR Lyrae stars given
in F98. They derived $M_{\rm V}$ = 0.65 $\pm$ 0.23 at an average
metallicity of [Fe/H] = $-1.51$.

The results by F98, T98 and L98 imply dimmer RR Lyrae stars by about
0.3 mag with respect to the results from either the Main Sequence
fitting technique using \HP\ subdwarfs (see, e.g., Gratton et
al. 1997) or recent theoretical Horizontal Branch models (see, e.g.,
Salaris \& Weiss 1998, Caloi et al. 1997). On the other hand, F98, T98
and L98 agree with the results from Baade-Wesselink analyses, which
predict a zero point of about 1.00 mag for the $M_{\rm V}$-[Fe/H]
relation (see, e.g., Clementini et al. 1995).

Turon Lacarrieu \& Cr\'ez\'e (1977) presented two methods to derive
the absolute magnitude of stars from the observed parallaxes, namely 
using individual LK-corrections and the method of ``reduced
parallaxes'' (hereafter RP) on a sample of stars. 
The advantages of the RP method are the following:
it avoids the biases due to the asymmetry of the errors when
transforming the parallaxes into magnitudes,
it can be applied to samples which contain negative parallaxes, it 
is free from LK-type bias if no selection on parallax, or error on the
parallax is made (Koen \& Laney 1998), and it requires no knowledge
about the space distribution of stars. We will apply the RP method to
the sample of 144 RR Lyrae stars used by F98, and
will derive the zero points of the $M_{\rm V}$-[Fe/H] and $M_{\rm
K}$-$\log P_{0}$ relations. Recently, Koen \& Laney (1998) also
briefly discussed the application of the RP method to RR Lyrae stars.
%

\vspace{-2mm}

\begin{table*}
\caption[]{Values for the zero point of the $M_{\rm V}$-[Fe/H] and
$M_{\rm K}$-$\log P_{0}$ relations from the RP method}
\begin{tabular}{crcrrrcrrl} \hline
        &   &  $M_{\rm V}$-[Fe/H] &   &       &    &  $M_{\rm
K}$-$\log P_{0}$ & &     &   \\                 
Solution& N & Zero point & Total  & Slope &  N &  Zero Point & Total & Slope &  Remarks \\
        &   &           & Weight &        &    &             & Weight &      &  \\ \hline
1 & 144 & 0.67 $\pm$ 0.24 & 45.5 & 0.18 &  108 &  $-$1.28 $\pm$ 0.25 & 241.3  &  $-$2.33  &  whole sample \\
2 &  84 & 0.77 $\pm$ 0.26 & 35.3 & 0.18 &  62 & $-$1.16 $\pm$ 0.27 &  188.4 & $-$2.33  &  [Fe/H] $\le -1.3$ \\
3 &  84 & 0.81 $\pm$ 0.24 & 40.0 & 0.18 &  62 & $-$1.14 $\pm$ 0.26 &
 201.4 & $-2.33$ &  [Fe/H] $\le -1.3$, ${\sigma}_{\rm H} = 0$ \\
4 & 144 & 0.64 $\pm$ 0.24 & 46.7 & 0.18 & &   &     &    & as 1, all [Fe/H] larger by 0.15 dex\\
5 & 144 & 0.60 $\pm$ 0.24 & 48.2 & 0.18 & 108 & $-$1.28 $\pm$ 0.25 & 242.8 & $-$2.33 &as 1, all $E(B-V)$ larger by 0.02 \\
\hline
\end{tabular}
\vspace{-3mm}
\end{table*}

\section{The ``reduced parallax'' method}

Let us consider a relation of the form:
\begin{equation}
 M_{\rm V} = \delta \,\, {\rm [Fe/H]} + \rho.
\end{equation}
If $V$ is the intensity-mean visual magnitude and $V_0$ its
reddening corrected value, then one can write:
\begin{equation}
10^{0.2\rho} = \pi \times 0.01\,\;10^{0.2(V_0 - \delta  \;{\rm
[Fe/H]} )} \equiv \pi \times {\rm RHS},
\end{equation}
which defines the quantity {\sc rhs} and where $\pi$ is the parallax
in milli-arcseconds. 
A weighted-mean of the quantity 10$^{0.2 \rho}$ is calculated, with the weight
(weight = $\frac{1}{{\sigma}^2}$) for the individual stars derived
from:
\begin{equation}
{\sigma}^2 = \left( {\sigma}_{\pi} \times {\rm RHS} \right)^2 + 
\left(0.2\,\ln(10) \,\; \pi\; {\sigma}_{\rm H}  \times {\rm RHS} \right)^2,
\end{equation}
with ${\sigma}_{\pi}$ the standard error in the parallax. This follows
from the propagation-of-errors in Eq.(2). 
We have adopted the slope $\delta = 0.18$ (see the discussion
in Fernley et al. 1998b), which is the one used by F98 and which is in
agreement with the results from Baade-Wesselink methods (see, e.g.,
Clementini et al. 1995), Main Sequence fitting (Gratton et al. 1997)
and theoretical models (see, e.g., Salaris \& Weiss 1998, Cassisi et
al. 1999).

The sample we consider is identical to that of F98, that is 144 stars
out of a total of 180 stars in the \HP\ catalogue. F98 discuss the
reasons for discarding the 36 stars.  Arguments include the fact that
these stars do not have reddening determinations, are not RR Lyrae
variables, or have poor quality \HP\ solutions.
Table~1 of F98 (retrievable from the CDS) lists all necessary data to
perform the above analysis: periods, intensity-mean $V$ and $K$
magnitudes, colour-excesses $E(B-V)$, and metallicities [Fe/H]. The
extinction is calculated from $A_{\rm V} = 3.1 E(B-V)$ (as done by F98).

An important requirement when applying this method is that the value
of ${\sigma}_{\rm H}$ is small compared to the errors on the parallax.
If the dispersion ${\sigma}_{\rm H}$ of the exponent in the factor RHS
is
large, the distribution of errors on the right-hand term in equation 2
is asymmetrical and a bias towards brighter magnitudes is introduced
(Feast \& Catchpole 1997, Pont 1999).  The adopted value of
${\sigma}_{\rm H}$ has been computed by considering four different
contributions: errors on the intensity-mean $V$ values of the RR Lyrae
stars (as given in Table~1 of F98), on the extinction (as derived from
the errors on E(B-V) given in Table~1 of F98), on [Fe/H] (again, from
Table~1 of F98), and the intrinsic scatter due to evolutionary effects
in the instability strip.  This last term is the most important one,
and we have adopted for it a 1$\sigma$ value by 0.12 mag (as in
Fernley et al. 1998b), following the results of the exhaustive
observational analysis by Sandage (1990). The final value is
${\sigma}_{\rm H}$ = 0.15, a quantity small enough in comparison with
the parallax errors so that no substantial bias is introduced on the
right-hand term of equation 2, as we have verified by means of
numerical simulations.  Even a ${\sigma}_{\rm H}$ of 0.20
mag. would lead to a bias by at most 0.02 mag.


Table~1 lists the values of the zero point with error we obtain with
different sample selections for the $M_{\rm V}$-[Fe/H] relation.
Solution 1 corresponds to the case of the whole sample; the zero point
of 0.67 $\pm$ 0.24 mag is about 0.4 mag brighter than the value
derived by F98, and consistent with the value listed in Koen \& Laney
(1998) using the same method with slightly different values for
${\sigma}_{\rm H}$. The sample with [Fe/H] $\le -1.3$ (Solution 2)
corresponds to a sample constituted entirely (according to the
discussion in F98) by Halo RR Lyrae stars, with a negligible
contamination from the Disk population. In this case the zero point is
equal to 0.77 $\pm$ 0.26 mag; it is slightly fainter than Solution 1,
but well in agreement within the statistical errors.  We also
re-derived the zero point for Solution 2 in the case of ${\sigma}_{\rm
H}$ = 0.0, and we found a change by only 0.04 mag.  A systematic
change in the metallicity scale (Solution 4) by 0.15 dex does not
affect appreciably the zero point determination, while the result is
more sensitive to a systematic variation of the adopted reddenings
(Solution 5).
%

The RP method has also been used to derive the zero point of the
$M_{\rm K}$-$\log P_{0}$ relation.  This relation appears to be
insensitive to the metallicity (Fernley et al. 1987, Carney et
al. 1995) and is also very weakly affected by reddening uncertainties,
since $A_{\rm K} = 0.112\, A_{\rm V}$ (Rieke \& Lebofsky
1985). Moreover, the intrinsic scatter around this relation is smaller
than in the case of the $M_{\rm V}$-[Fe/H] relation (Fernley et
al. 1987). In the sample considered here there are 108 RR Lyrae stars
with an observed intensity-mean $K$ magnitude.  The procedure is the
same as described before, the only difference is that now, instead of
Eq.~1, we use the expression $ M_{\rm K} = \delta \,\, {\rm \log
P_{0}} + \rho$ where $P_{0}$ is the fundamental pulsation period.  For
the first-overtone RRc variables we have derived the fundamental
periods using the relation $\log P_{0}/P_{1}$ = +0.120 (Carney et
al. 1995).  We adopt a slope $\delta = -2.33$ following Carney et
al. (1995); for the value of ${\sigma}_{\rm H}$ we have considered the
same contributions previously described (with the exception, of
course, of the contribution due to the error on [Fe/H]). In this case
the observational estimate of the intrinsic scatter due to the width
of the instability strip comes from Carney et al. (1995), and the
final value results to be ${\sigma}_{\rm H}$=0.10.

In Tab.~1 the values of the zero point for the $M_{\rm K}$-$\log
P_{0}$ relation are listed.  When considering the entire sample we
obtain a zero point of $-1.28 \pm 0.25$ mag, $\approx$ 0.4 mag
brighter than the value from the Baade-Wesselink method (see, e.g.,
Carney et al. 1995). In the case of a pure Halo RR Lyrae sample
([Fe/H]$ \leq -1.3$) we obtain $-1.16 \pm 0.27$ mag, slightly dimmer
but again in agreement with the value derived for the whole sample.
The influence of ${\sigma}_{\rm H}$ is even less than for the $M_{\rm
V}$-[Fe/H] relation.
%

As the sample of the RR Lyrae stars is not volume complete it may be
subject to Malmquist type bias. If the space distribution of RR Lyrae
is spherical it implies that the true zero points of the $M_{\rm
K}$-[Fe/H] and $M_{\rm K}$-$\log P_{0}$ relations may be fainter by up
to 0.03 and 0.01 mag, respectively, for the adopted values of
${\sigma}_{\rm H}$. This applies when average absolute magnitudes of a
volume and brightness limited sample are compared. Oudmaijer et
al. (1999) showed empirically that when the averaging is done over
10$^{0.2M_{\rm V}}$ the effect of Malmquist bias is less. \\

\begin{figure}
\centerline{\psfig{figure=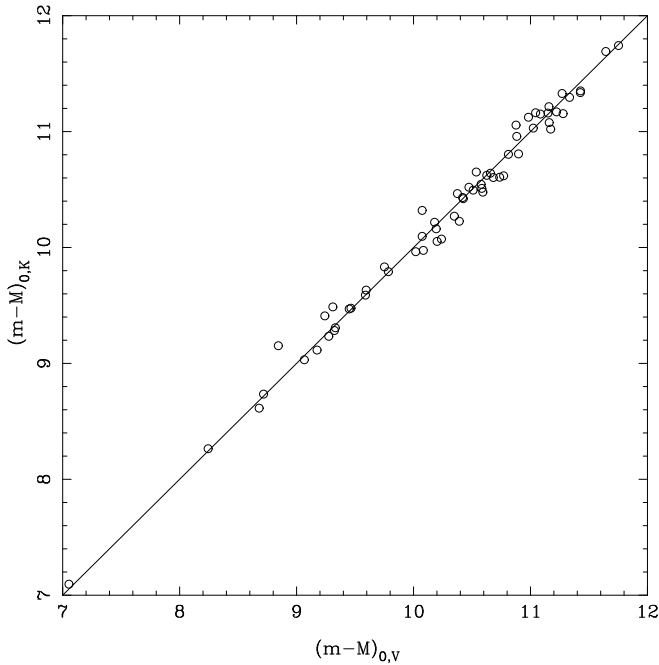,width=8.8cm,angle=-90}}
\caption[]{A comparison of the true distance moduli to the 62
metal-poor RR Lyrae with [Fe/H] $\le -1.3$ from the $M_{\rm}-$[Fe/H]
and $M_{\rm K}-\log P$ relations. Each point has an error bar of about
0.26 mag in both $x-$ and $y-$direction. The solid line is the
1:1 relation. The dispersion is less than 0.10 mag.}
\vspace{-3mm}
\end{figure}

\noindent

In Fig.~1 we compare, for the 62 \HP\ RR Lyrae stars with [Fe/H] $\le
-1.3$ and both observed K and V magnitudes, the true distance moduli
derived from the $M_{\rm V}$-[Fe/H] and $M_{\rm K}$-$\log P_{0}$
relations, using zero points of $0.77$ and $-1.16$ mag, respectively.
Each data point has an error bar of 0.26 mag in $x-$ and 0.27 mag in
the $y-$direction.
The comparison of the two photometric distances can in principle give
us an independent indication for possible biases in the determination
of the zero points of the two
relations with the RP method.  As it is evident from the figure, the
distance moduli from both relations agree very well.  A linear fit to
the data is consistent with a slope of unity, and
%
%
the dispersion around the 1:1 relation is equal to 0.098 mag. 
A dispersion of this order is what is expected from the
dispersions in the observed $\log P$-[Fe/H] and $(V-K)_0-\log P$
relations for the RR Lyrae sample.

\begin{table*}
\caption[]{Data on RR Lyrae in LMC clusters}
\begin{tabular}{cccrrrcc} \hline
  Name & $<V>$ & $A_{\rm V}$ & [Fe/H] & $\Delta$ & Reference & $V_{0}
+\Delta$ &$V_{0} +\Delta -$0.18[Fe/H] \\
 & (mag.) & (mag.)  & & (mag.) & &(mag.)  &(mag.)  \\ \hline 

NGC 1466 & 19.33 $\pm$ 0.02 & 0.28 $\pm$ 0.06 & $-1.85$ & 0.0 & Walker 1992b 
& 19.05 $\pm$ 0.07 & 19.38 $\pm$ 0.07 \\ 
NGC 1786 & 19.27 $\pm$ 0.03 & 0.23 $\pm$ 0.03 & $-2.3$ & 0.0 & 
Walker \& Mack 1988 & 19.04 $\pm$ 0.04 & 19.45 $\pm$ 0.04 \\ 
NGC 1835 & 19.38 $\pm$ 0.05 & 0.40 $\pm$ 0.09 & $-1.8$ & $-0.03$ &
Walker 1993 & 18.94 $\pm$ 0.11 & 19.26 $\pm$ 0.11 \\ 
NGC 2210 & 19.12 $\pm$ 0.10 & 0.19 $\pm$ 0.09 & $-1.9$ & $+0.09$ & 
Reid \& Freedman 1994 & 19.02 $\pm$ 0.13 & 19.36 $\pm$ 0.13 \\ 
Reticulum & 19.07 $\pm$ 0.03 & 0.09 $\pm$ 0.06 & $-1.7$ & $-0.08$ & 
Walker 1992a & 18.91 $\pm$ 0.07 & 19.22 $\pm$ 0.07 \\ 
NGC 1841 & 19.31 $\pm$ 0.02 & 0.56 $\pm$ 0.06 & $-2.2$ & ($\sim$ 0.2)
& Walker 1990 & & \\  
NGC 2257 & 19.03 $\pm$ 0.02 & 0.12 $\pm$ 0.03 & $-1.8$ & +0.18 & 
Walker 1989 & & \\ 
\hline
\end{tabular}
\vspace{-3mm}
\end{table*}

\section{Discussion}

For their preferred sample of 84 stars with [Fe/H] $\le -1.3$ F98
obtain a zero point of 1.05 $\pm$ 0.15 mag for the $M_{\rm
V}$-[Fe/H] relation (assuming a slope of 0.18), in agreement with
results from Baade-Wesselink methods. When applying the RP method to
the same sample of stars, we find a zero point 0.28 mag brighter. An
analogous result, which means a zero point $\approx$0.30 mag brighter
than the Baade-Wesselink one, is derived for the $M_{\rm K}$-$\log
P_{0}$ relation.

Even if within the error bar the results derived with the different
methods formally agree, there appears to exist a systematic difference
between zero points obtained using the parallaxes directly and zero
points obtained by employing methods which are sensitive to proper
motions and radial velocities (F98, T98, L98), especially if one also
takes into account the results for the \HP\ Cepheids.  Also with the
\C\ one finds that methods where the results are mostly sensitive to
the proper motions and radial velocities find dimmer zero points for
the \C\ PL-relation compared to methods which directly use the
parallax. In particular, using the RP method Feast \& Catchpole (1997)
derived a zero point of $-1.43 \pm 0.10$ mag, and Lanoix et al. (1999)
using a slightly bigger sample find $-1.44 \pm 0.05$ mag. Oudmaijer et
al. (1998), using only the positive parallaxes but then correcting for
the LK-bias, find $-1.29 \pm 0.08$ mag. On the other hand, L98 find a
zero point of $-1.05 \pm 0.17$ mag using a maximum likelihood method
that takes into account parallaxes, proper motions and velocity
informations. As discussed by Pont (1999), in this technique the
parallaxes do not influence the result to first order, and the method
is similar to a statistical parallax analysis.  A careful check of all
assumptions implicit in the kinematical methods could be the key to
understanding the nature of this puzzling disagreement. In the case of
the RP method, as discussed extensively in the previous section, the
condition for deriving the zero point without introducing a bias is to
have ${\sigma}_{\rm H}$ small with respect to the errors on the
parallaxes; this condition appears to be fulfilled in the sample considered.

Our zero point for the $M_{\rm V}$-[Fe/H] relation is in agreement
with results from the Main Sequence fitting technique (Gratton et
al. 1997), and from theoretical Horizontal Branch models.  In
particular, the Horizontal Branch models by Salaris \& Weiss (1998)
and Cassisi et al. (1999) give a zero point for the Zero Age
Horizontal Branch (ZAHB) at the RR Lyrae instability strip in the
range 0.74-0.77 mag. To compare the results for the ZAHB brightness
with the $M_{\rm V}$-[Fe/H] relations mentioned in this paper which
consider the mean absolute brightness of the RR Lyrae stars population
at a certain metallicity, one has to apply a correction by $\approx
-0.1$ mag (see, e.g., Caloi et al. 1997 and references therein) to the
ZAHB result; this takes into account the evolution off the ZAHB of the
observed RR Lyrae stars.  Even after applying this correction the
theoretical results are in good agreement with the results from the RP
method.  Moreover, the zero point derived with the RP method is also
in agreement with the recent results by Kovacs \& Walker (1999), who
derive, by employing linear pulsation models, RR Lyrae luminosities
that are brighter by 0.2-0.3 mag with respect to Baade-Wesselink results.

Finally, we want to derive the LMC distance implied by our zero point
of the RR Lyrae distance scale. 
Table~2 collects the available data on RR Lyrae stars in LMC clusters:
the name of the cluster, the observed mean $V$-magnitude, reddening,
metallicity and the difference in distance modulus ($\Delta$) between
the cluster and the main body of the LMC. All these data are taken
from the references listed. From them the dereddened magnitude at the
centre of the LMC (Col.~7), and this value minus the quantity (0.18
[Fe/H]) (Col.~8) have been calculated for those clusters with $\Delta
<$0.1 mag.  At this point we have taken into account the difference in
metallicity between the clusters before deriving the LMC distance. 
More in detail, we have derived the weighted mean of the values in
Col.~8 to find an average of 19.38 with a rms dispersion of 0.10 mag,
which can be compared directly to the zero point of the $M_{\rm
V}-$[Fe/H] relation to find a distance modulus of 18.61 $\pm$ 0.28.
This result turns out to be consistent with the \C\ distance to the
LMC as derived by Feast \& Catchpole (1997) or Oudmaijer et al. (1998).

\vspace{-5mm}

\subsection*{Acknowledgements}
Ren\'e Oudmaijer, Phil James and the referee, Xavier Luri,
are  warmly thanked for valuable comments and suggestions which
improved the presentation of the paper.
This research has made use of the SIMBAD database,
operated at CDS, Strasbourg, France.

{}

\end{document}